# A generic rule that simplifies the derivation of the transformation equations accounting for the properties of the photon

**Bernhard Rothenstein[1] and Stefan Popescu[2]**
1) Politehnica University of Timisoara, Physics Department,
Timisoara, Romania brothenstein@gmail.com
2) Siemens AG, Erlangen, Germany stefan.popescu@siemens.com

***Abstract***. *We show that the transformation equation for the tardyon velocity involves two generic functions which in turn depend on the relative velocity of the involved reference frames, on the tardyon velocity u and on the polar angle which define the direction along which the tardyon moves. The same functions are further involved in the transformation equations for the space-time coordinates of the same event generated by a moving tardyon and for its relativistic mass, momentum and energy. Taking the limits of these functions for u approaching c we obtain exactly the transformation equations for the space-time coordinates of the same event generated by a photon and for its momentum and energy. The same procedure works also for the transition from a plane acoustic wave to an electromagnetic wave.*

## 1. Introduction

The event is a fundamental concept in physics defined as something that takes place at a given point in space at a given time. In order to define the point where the event takes place in a two space dimensions approach we introduce an inertial reference frame K(XOY) and let (x,y) and (r,θ) be the Cartesian and the polar space coordinates of this point expressed by

$$x = r\cos\theta \qquad (1)$$
$$y = r\sin\theta. \qquad (2)$$

Having also a clock located at this point that reads *t* when the event takes place we are able to completely define the event by its space-time coordinates $E(x = r\cos\theta; y = r\sin\theta, t)$. In order to be operational all the clocks located at different points in K(XOY) should display the same running time t, a situation that could be achieved by the clock synchronization procedure proposed by Einstein[1].

Special relativity becomes involved when we consider a second inertial reference frame K'(X'O'Y'). The corresponding axes of the two frames are parallel to each other, the axes OX and O'X' are overlapped and at the origin of time in the two frames (t=t'=0) the origins O and O' are located at the same point in space. The frame K' moves with constant velocity V relative to K in the positive direction of the overlapped axes. Event E detected from K' is defined by the space-time coordinates $E'(x' = r'\cos\theta', y' = r'\sin\theta', t')$. One of the fundamental problems in special relativity is to find out the relationships between the space-time coordinates of events E and E' defined above in accordance with Einstein's relativistic postulate.

Events are generated by moving physical entities and special relativity operates with two kinds of entities:

- **tardyons** which move with different subluminal velocities u<c relative to different reference frames in relative motion in respect to each other. We define the particular inertial reference frame relative to which the tardyon is in a state of rest as being its proper inertial reference frame.

- **photons (light signals)** which move with the same speed **c** relative to any inertial reference frames and we could not identify a rest frame for any of it.

We characterize the tardyon by its **proper mass** measured by observers relative to whom it is in a state of rest and by its **relativistic mass** measured by observers relative to whom it moves. The **tardyon** is additionally characterized by its **momentum, proper energy, kinetic energy** and **total energy.** The photon generates events as well and we characterize it by its **relativistic mass, momentum** and **energy** but not by a proper mass because the photon exists only in state of motion. We also associate a special clock with the moving tardyon. This can be the wrist watch of the observer moving together with the tardyon. Equipped with the wrist watch an observer measures **a proper time interval** as a difference between the time coordinates of two events that take place in front of him. Two observers located at different points and working in team measure a **non-proper time interval** as the difference between the readings of theirs clocks when two different events take place in front of each respectively. Considering a rod we characterize it by its **proper length** measured by observers relative to whom it is in a state of rest and by its **non-proper length** measured by observers relative to whom it moves.

The **transformation equations** establish a relationship between physical quantities measured by observers from K and K' respectively. We know transformation equations which relate the proper magnitude of a physical quantity measured in one of the reference frames to its non-proper magnitude measured in the other one. As a typical example we mention the formula that accounts for the **time dilation relativistic effect.**

The purpose of our paper is to show that knowing a transformation equation for physical quantity characterizing some property of the **tardyon** we can obtain the transformation equation for the same physical quantity characterizing a **photon** by simply replacing in the transformation formulas the speed of the tardyon (**u** in K and **u'** in K') with **c.**

## 2. Deriving the transformation equations that account for the tardyon behavior

The scenario we follow involves a tardyon start moving at t=t'=0 from the overlapped origins O and O' of the involved reference frames. The line along which the tardyon moves makes the angle θ with the positive direction of the overlapped axes OX and O'X'. The tardyon moves with speed $\mathbf{u}(u_x, u_y)$ relative to K and $\mathbf{u}'(u'_x, u'_y)$ relative to K'. After a given time of motion $t(t')$ the tardyon generates the event $E(x = r\cos\theta; y = r\sin\theta, t = r/u)$ which when detected from K' is characterized by the space-time coordinates $\mathbf{E}'(x' = r'\cos\theta'; y' = r'\sin\theta'; t' = r'/u')$.

The tardyon is characterized by a rest mass $m_0$, relativistic mass m', momentum p', rest energy $E_0$ and a total energy E' in K' and by the corresponding physical quantities m, p, E in K. It is well known that the addition law of velocities can be derived without using the Lorentz-Einstein transformations (LET) for the space-time coordinates of the same event[1]. We present them as

$$u_x = u' \frac{\cos\theta' + V/u'}{1 + \frac{Vu'}{c^2}\cos\theta'} = \frac{\psi_1(V, u', \theta')}{\psi_2(c, V, u', \theta')} \qquad (3)$$

and

$$u_y = u' \frac{\sqrt{1 - V^2/c^2} \sin\theta'}{1 + \frac{Vu'}{c^2}\cos\theta'} = u' \frac{\gamma^{-1}\sin\theta'}{\psi_2(c, V, u', \theta')} \qquad (4)$$

the magnitudes transforming as

$$u = u' \frac{\sqrt{\psi_1^2 + (1 - V^2/c^2)\sin^2\theta'}}{1 + \frac{Vu'}{c^2}\cos\theta'} = \frac{\psi_3(c, V, u', \theta')}{\psi_2(c, V, u', \theta')} \qquad (5)$$

The generic functions $\psi_{1-3}$ are given by:

$$\psi_1 = \cos\theta' + V/u'; \quad \psi_2 = 1 + V\frac{u'}{c^2}\cos\theta' \text{ and } \psi_3 = \sqrt{\psi_1^2 + V\frac{u'}{c^2}\cos\theta'}$$

Consider an extra clock $C_0^0$ that moves with speed $\mathbf{u}'(u'_x, u'_y)$ relative to K' and with speed $\mathbf{u}(u_x, u_y)$ relative to K. This clock reads $t^0 = 0$ when located in front of clocks $C_0(0,0)$ and $C'_0(0,0)$ of the reference frames K and K' which also read t=t'=0. In the scenario we follow after a given time of motion when clock $C_0^0$ reads $t^0$ it is located in front of a clock $C(x, y)$ of the K frame which reads t and in front of a clock $C'(x', y')$ of the K' frame which reads t'.

In accordance with the **time dilation** relativistic effect[2] the readings of the three clocks when they are located at the same point in space are related by

$$t = \frac{t^0}{\sqrt{1-\frac{u^2}{c^2}}} \quad (6)$$

$$t' = \frac{t^0}{\sqrt{1-\frac{u'^2}{c^2}}}. \quad (7)$$

Taking into account the conditions imposed to the origins of space and time in K and K' we conclude that (6) and (7) hold also for the time intervals measured by the mentioned clocks $\Delta t^0, \Delta t$ and $\Delta t'$. We mention that $\Delta t^0$ represents a proper time interval whilst $\Delta t; \Delta t'$ represent non-proper (coordinate) time intervals. Expressing the right side of (6) as a function of physical quantities measured by observers of frame K' via (5) and taking into account (6) we obtain that

$$t = \frac{\psi_2}{\sqrt{1-V^2/c^2}} t' \quad (8)$$

which represents the LET for the time coordinates of the same event. By definition

$$r = ut \quad (9)$$

and

$$r' = u't' \quad (10)$$

represent the magnitudes of the vectors r and r' tracking the position of clock $C_0^0$ in frame K and K' respectively. Expressing the right side of (9) as a function of physical quantities measured from K' via (5) and (8) we obtain

$$r = r' \frac{\sqrt{\psi_1^2 + (1-V^2/c^2)\sin^2\theta'}}{\sqrt{1-V^2/c^2}} = \frac{\psi_3(c,V,u',\theta')}{\sqrt{1-V^2/c^2}} \quad (11)$$

Taking into account the invariance of distances measured perpendicular to the direction of **V** (a direct consequence of the principle of relativity[3]) we have

$$r \sin\theta = r \sin\theta' \quad (12)$$

and so the angles $\theta$ and $\theta'$ which define the directions along which the tardyon moves when detected from K and K' respectively are related by

$$\sin\theta = \frac{\sqrt{1-V^2/c^2}}{\psi_3} \sin\theta' \quad (13)$$

With some simple trigonometry we can derive the transformation equations for the other trigonometric functions of θ and θ'. The distance x traveled by clock $C_0^0$ in the positive direction of the overlapped axes when detected from K and x' when detected from K' transform as

$$x = x' \frac{\psi_1}{\sqrt{1-V^2/c^2}}. \tag{14}$$

Let $m_0$ be the rest mass of clock $C_0^0$. When moving with speed **u** relative to K its mass in K is

$$m = \frac{m_0}{\sqrt{1-u^2/c^2}} \tag{15}$$

whereas when moving with **u'** relative to K' its mass in K' is

$$m' = \frac{m_0}{\sqrt{1-u'^2/c^2}}. \tag{16}$$

We consider that (15) and (16) are the result of bringing experimental results in accordance with special relativity theory[4]. The similitude with (6) and (7) leads directly to the following relationship between the relativistic (non-proper) masses m and m'

$$m = m' \frac{\psi_3}{\sqrt{1-V^2/c^2}} \tag{17}$$

Multiplying both sides of (17) with $c^2$ we obtain that the energy of the tardyon transforms as

$$E = E' \frac{\psi_3}{\sqrt{1-V^2/c^2}}. \tag{18}$$

The momentum of the tardyon as detected from K is by definition

$$p = mu \tag{19}$$

whereas when detected from K' it is

$$p' = m'u'. \tag{20}$$

Again analogy with the previous case leads to

$$p = p' \frac{\psi_3}{\sqrt{1-V^2/c^2}}. \tag{21}$$

Taking into account that the position vector, velocity and momentum are collinear the involved angles transform in accordance with (13) whereas the components of the momentum transform as

$$p_x = p'_x \frac{\psi_1}{\sqrt{1-V^2/c^2}} \tag{22}$$

$$p_y = p'_y. \tag{23}$$

The approach to relativistic dynamics presented above avoids the involvement of conservation laws saving the time allocated to teach the special relativity.

**3. Deriving the transformation equations that account for the photon behavior by a simple generic rule**

In the scenario above we replace the tardyon by a photon characterized by its inertial mass $m'_c$, momentum $p'_c$ and energy $E'_c$ in K' and respectively by $m_c$, $p_c$ and $E_c$ in K. The photon moves with the same speed **c** relative to both frames and generates after a given time of motion $t(t')$ the event $E_c(x_c = r_c \cos\theta_c; y_c = r_c \sin\theta_c; t_c = r_c/c)$ as detected from K and $E'_c(x'_c = r'_c \cos\theta'_c; y'_c = r'_c \sin\theta'_c; t'_c = r'_c/c)$ as detected from K'.

Consider the following rule: **Knowing the transformation equations for the space-time coordinates of the same event generated by a tardyon and for its relativistic mass, energy and momentum we obtain the corresponding equations for a photon by indexing the physical quantities with c and taking the limit with the tardyon velocity approaching the light speed.** Following this simple rule we obtain the transformation equations for the space time coordinates of the same event generated by the photon as

$$t_c = \gamma \psi_{1,c} \tag{24}$$

$$r_c = \gamma \psi_{3,c} r'_c \tag{25}$$

$$\sin\theta_c = \gamma^{-1} \frac{\sin\theta'_c}{\psi_{3,c}}. \tag{26}$$

Furthermore we obtain the equations which define the dynamical behavior of the photon. The relativistic mass transform as

$$m_c = \gamma m'_c \psi_{1,c} \tag{27}$$

the energy as

$$E_c = \gamma E'_c \psi_{1,c} \tag{28}$$

and the magnitude and components of momentum as

$$p_c = \gamma p'_c \psi_{3,c} \tag{29}$$

$$p_{c,x} = \gamma p'_{c,x} \psi_{1,c} \tag{30}$$

$$p'_{c,y} = p_{c,y} \tag{31}$$

These results are in good accordance with those obtained by laborious derivations that we find in the literature of the subject.

Moller[5] showed that the frequency of the oscillations taking place in an acoustic wave transforms as

$$f = \gamma f' \psi_3. \tag{32}$$

In accordance with our generic rule the frequency of the electromagnetic oscillations taking place in the electromagnetic wave and the frequency of the photon associated to it transforms as

$$f_c = \gamma f'_c \psi_{3,c}. \tag{33}$$

The $\psi$ functions with index c introduced above are given by

$$\psi_{1,c} = \lim_{u' \to c} \psi_1 = \cos\theta'_c + V/c \tag{34}$$

$$\psi_{2,c} = \lim_{u' \to c} \psi_2 = 1 + \frac{V}{c}\cos\theta'_c \tag{35}$$

$$\psi_{3,c} = \lim_{u' \to c} \psi_3 = 1 + \frac{V}{c}\cos\theta'_c = \psi_{2,c} \tag{36}$$

and that justifies the generic rule never meaning that the tardyon could become a photon.

## 4. Electric and magnetic field. Electromagnetic wave

Consider a linear distribution of charge located along the overlapped axes OX(O'X') that moves with speed $u_x$ relative to K and with speed $u'_x$ relative to K'. This charge generates at a point $M^0(0,d)$ in its rest frame an electrostatic field

$$E^0_y = \frac{\lambda^0}{2\pi\varepsilon_0 d} \tag{37}$$

$\lambda_0$ representing the linear density of charge.[6] and we consider only its component which shows in the positive direction of the OY axis in a two space dimensions approach. Detected from K the same electric field is given by

$$E_y = \frac{\lambda^0}{2\pi\varepsilon_0 d\sqrt{1-u_x^2/c^2}} \tag{38}$$

whereas detected from K' it is

$$E'_y = \frac{\lambda^0}{2\pi\varepsilon_0 d\sqrt{1-u'^2_x/c^2}}. \tag{39}$$

The relativistic addition law of speeds leads to the following relationship between $E_y$ and $E'_y$

$$E_y = E'_y \frac{1+Vu'_x/c^2}{\sqrt{1-V^2/c^2}} = \frac{E'_y + V\frac{u'_x}{c^2}E'_y}{\sqrt{1-V^2/c^2}} = \frac{E' + VB'_z}{\sqrt{1-V^2/c^2}}. \tag{40}$$

using the notation $B'_z = u'_x E'_y/c^2$ for the O'Z' component of the magnetic field. In K it reads $B_z = u_x E_y/c^2$ and transforms as

$$B_z = B'_z \frac{1+V/u'_x}{\sqrt{1-V^2/c^2}} = \frac{B'_z + VE'_y/c^2}{\sqrt{1-V^2/c^2}} \tag{41}$$

As we see the same functions $\psi_1$ and $\psi_2$ are involved in the transformation process as in the previous cases. The vectors $(E_y, B_z)$ and $(E'_y, B'_z)$ are perpendicular to each other and so are they in the electromagnetic wave that propagates with speed c in the positive direction of the OX(O'X') axes. In accordance with our rule they transform in this case as

$$E_{y,c} = E'_{y,c} \sqrt{\frac{1+V/c}{1-V/c}} \qquad (42)$$

and

$$B_{z,c} = B'_{z,c} \sqrt{\frac{1+V/c}{1-V/c}} \qquad (43)$$

respectively.

### 4. Conclusions

We have shown that the transformation equation for the tardyon velocity defines three generic functions, that depend on the relative velocity of the involved inertial reference frames, tardyon velocity $u$ and polar angle which define the direction along which the tardyon moves. The same functions are further involved in the transformation equations for the space-time coordinates of the same event generated by a moving tardyon and for its relativistic mass, momentum and energy. Taking the limits of these generic functions for $u \to c$ we obtain exactly the transformation equations for the space-time coordinates of the same event generated by a photon and for its momentum and energy. Our method works also for the electromagnetic wave.